\documentclass[12pt]{article}

\usepackage[greek,english]{babel}
\usepackage{graphicx}
\usepackage{caption}
\usepackage{lscape}
\usepackage{rotating}
\usepackage{epstopdf}
\usepackage{algorithm}
\usepackage{color}
\usepackage[authoryear]{natbib}

\input{epsf}

\newcounter{saveeqn}

\newcommand{\bfpsi}{\mbox{\boldmath $\psi$}}

\newcommand{\bfphi}{\mbox{\boldmath $\phi$}}

\newcommand{\bfz}{\mbox{\boldmath $z$}}

\newcommand{\bfx}{\mbox{\boldmath $x$}}

\begin{document}

\parindent0mm \parskip0.6cm

{\LARGE {\bf  On synthetic interval data with predetermined subject partitioning, and partial control of the variables' marginal correlation structure} }

\begin{large}

Michail Papathomas \footnote{Michail Papathomas (corresponding author) is a Senior Lecturer in Statistics, School of Mathematics and Statistics,  University of St Andrews, The Observatory,  Buchanan Gardens, St Andrews, KY16 9LZ, UK (e-mail:M.Papathomas@st-andrews.ac.uk)}

\end{large}

{\it School of Mathematics and Statistics, University of St Andrews, United Kingdom }

\newpage

{\bf ABSTRACT.} 

A standard approach for assessing the performance of partition models is to create synthetic data sets with a prespecified clustering structure, and assess how well the model reveals this structure. A common format is that subjects are assigned to different clusters, with observations simulated so that subjects within the same cluster have similar profiles, allowing for some variability. In this manuscript, we consider observations from interval variables, taking a finite number of values. Interval data are commonly observed in cohort and Genome Wide Association studies, and our focus is on Single Nucleotide Polymorphisms. Theoretical and empirical results are utilized to explore the dependence structure between the variables, in relation with the clustering structure for the subjects. A novel algorithm is proposed that allows to control the marginal stratified correlation structure of the variables, specifying exact correlation values within groups of variables. Practical examples are shown, and a synthetic dataset is compared to a real one, to demonstrate similarities and differences.


\newpage


\section{Introduction}

Partitioning and mixture models are often used to reveal the clustering structure within a sample. For example, to discover if combinations of risk factors are associated with the risk of disease (M\"{u}ller et al. 2011), or to reveal dependencies in a population, whilst reducing the dimensionality of the problem; Yau \& Holmes (2011). In Bhattacharya and Dunson (2012) tensor factorizations are employed to characterize the joint density of variables that form high-dimensional data.  In Zhou et al. (2015) and Papathomas and Richardson (2016), marginally independent variables are detected with the use of modelling that is directly related to Bayesian partitioning algorithms. An overview of clustering approaches is given in Hennig et al. (2015) and Fr{\"u}hwirth-Schnatter (2016). 

We adopt a model based approach and define as cluster each one of the components of the adopted mixture model. Therefore, we undertake that two subjects belong to the same cluster when the corresponding vectors of observations are generated by the same component of the mixture model. See Hennig (2015) for an extensive discussion on cluster definition. Assessing the performance of partitioning models involves the creation of synthetic data with a prespecified clustering structure. The model is then fitted to the simulated data to evaluate its performance in terms of revealing this structure. Usually, profiles are created for a number of subjects, by simulating observations from a set of variables. The subjects are assigned to different clusters, and variable observations are simulated so that subjects within the same cluster have similar profiles, allowing for some variability. The investigator controls the strength of the signal in the clustering structure (i.e. how distinct the different clusters are), and the variability of the observations within each cluster. Sometimes partitioning the variables is also of interest (Marbac et al. 2014; Kirk et al. (2023)). In this manuscript we focus on the former more standard set-up, as the two frameworks can be viewed as interchangeable for simulated observations. 

Partitioning models for continuous observations (Jasra et al. 2005) often allow for a specific correlation structure for the variables, given the cluster allocation. This typically involves a multivariate normal distribution (Jing et al. 2024). In contrast to continuous observations, clustering approaches for observations from interval variables typically prescribe that variables are independent given the clustering of the subjects (Dunson and Xing 2009; Liverani et al. 2015). The resulting dimensionality reduction is the main advantage of this local independence modelling, as determining a fully specified joint distribution between $P$ variables with $M$ levels requires the specification of $M^{P}$ probabilities, a task that quickly becomes cumbersome and unwieldy. Celeux and Covaert (2016) comment on this notable modelling difference, mentioning that, in many applications, conditional independence has proven successful in achieving the main objective of a clustering algorithm, which is to bring together similar observations. In Oberski (2016), local dependence is discussed, given well defined substantive interest. In this manuscript, we concentrate on interval variables, and adopt the widely espoused independence assumption conditionally on the clustering of the subjects.  

In all examples we ensure that clustering structures are identifiable, up to label switching, by following the guidelines of Allman et al. (2009) for the required number of variables for mixture models where the within cluster independence assumption holds. Thus, denoting by $C$ the number of clusters, all synthetic datasets satisfy the identifiability condition, $P \geq 2[\mbox{log}_{M}(C)]+1$. 


Our work concerns interval variables, where the numerical distance between possible values is meaningful and known. Interval variables are of particular interest to us, as data from epidemiological and association cohort studies, such as number variants, are often in this form. Furthermore, continuous observations are often transformed to interval ones, when data from cohort studies are analyzed. This is done to alleviate the adverse effect of outlier observations (for example in dietary observations; see Bingham and Riboli 2004), or to allow for the flexible modelling of interactions (for example in air pollution variables; see Papathomas et al. 2011). Importantly, interval variables allow for the use of covariances and correlations through expectations. 

The variables are independent given the clustering of the subjects, but marginally dependent. In synthetic data sets, the induced marginal dependence can be at odds with the dependence structure observed in real data sets. 
The creation of synthetic data with predetermined clustering structure is straightforward, as long as the marginal dependence structure between the variables, generated as a by-product of the clustering structure, is ignored. 
In this manuscript, an algorithm is proposed where the clustering structure is predetermined, while maintaining partial control over the marginal dependence structure between the variables. This enables 
the creation of simulated data sets that share more characteristics with real ones, compared to synthetic data created with standard methods. To the best of our knowledge, no such algorithm has yet been proposed. 
Approaches in the literature relevant to marginal correlations focus on continuous observations, and on deriving association measures that provide unbiased estimates of marginal correlations when the size of the cluster relates to some outcome and a random-effects type model is utilized (Lorenz et al. 2011; Paulou et al. 2013). Consequently, the relevance of these approaches to the work presented in this manuscript is limited.  Wang and Sabo (2015) discuss the simulation of correlated binary observations, incorporating cluster specific random effects, but the aim of the proposed algorithm is not to generate clusters with distinct variable profiles. 

Our focus is on generating simulated data sets that contain observations that emulate Single Nucleotide Polymorphisms (SNP), although the proposed methods are more generally applicable. We do not touch on issues relevant to recombination and imputation (Ioannidis et al. 2009), as this is beyond the scope of this manuscript. 

In Section 2, we describe the generic approach for creating data with a predetermined clustering structure, and explore the marginal dependence structure between interval variables, deriving theoretical results. 
In Section 3, we 
introduce a specific algorithm for constructing clusters with distinct variable profiles and examine its properties. We focus on SNP-like simulated observations and derive results that effect control on the marginal dependence of the variables, in tandem with practical examples. In Section 4, a real data set containing SNP observations is compared to a synthetic one, demonstrating similarities and differences. We conclude with a discussion in Section 5.

\section{Simulating a predetermined clustering structure and the implied correlation matrix}

\subsection{The clustering model}

Assume $P$ variables $x_{.p}$, $p=1,\ldots ,P$. Without any loss of generality, assume that each variable takes values $1,\ldots ,M_{p}$. Let $\bfx=(x_{.1},\ldots ,x_{.P})$. Denote by $n$ the number of subjects. For subject $i$, $i=1,\ldots ,n$, a variable profile $x_{i}$ is a set of values $x_{i}=\{x_{i1},\ldots ,x_{iP} \}$. Let $\bfz=(z_1,\ldots ,z_n)$, where $z_{i}$ is an allocation variable, so that $z_{i}=c$ denotes that subject, $i$, belongs to cluster $c$. Denote by 
$\phi^{c}_{p}(x)$ the probability that $x_{.p}=x$, when the individual belongs to cluster $c$. Given the clustering allocation, the variables are assumed independent, each one following a multinomial distribution with cluster specific parameters
$\bfphi^{c}_{p}=(\phi^{c}_{p}(1),\ldots ,\phi^{c}_{p}(M_{p}))$. 
Denote by $\bfpsi = \{ \psi_{1}, \psi_{2}, \ldots , \psi_{C} \}$ the probabilities that a subject belongs to cluster $c$, $c=1, \ldots , C$. For more on finite Bayesian mixture models see Gr\"un and Malsiner-Walli (2022). 

\subsection{A generic algorithm for a predetermined clustering structure}

A generic algorithm for creating observations from $P$ variables, for subjects that are partitioned in $C$ clusters, is given as: 
\begin{itemize}
\item Specify the number of clusters $C$. 
\item Specify the number of subjects $n_{c}$, $c=1,\ldots ,C$, allocated to cluster $c$, in accordance with cluster allocation probabilities $\psi_c$.  Allocate subjects to clusters so that $n_c$ subjects exactly are allocated to cluster $c$. [Alternatively, a cluster can be drawn for a subject according to the allocation probabilities.]
\item Specify the variable profile of the subjects within each cluster, i.e. probabilities 
$P(x_{.p}=x_{ip} | z_{i}=c)=\phi_{p}^{c}(x_{ip})$, for all $c=1,\ldots ,C$, $p=1,\ldots ,P$, and $x_{ip}=1,\ldots ,M_{p}$, to generate a distinct variable profile for the subjects in each cluster.  
\item To generate $x_{ip}$, sample from a multinomial distribution with probabilities $\bfphi^{z_i}_{p}$.
\end{itemize}

\subsection{The marginal correlation structure of interval variables}

Assume that $\bfx$ is a vector of interval variables. The marginal variance-covariance matrix is, 
\[
\mbox{Var}(\bfx) = \mbox{E}(\bfx \bfx^{\top}) - \mbox{E}(\bfx)\mbox{E}(\bfx)^{\top}
\]
\[
=\mbox{E}_{\bfz} \mbox{E}_{\bfx|\bfz}(\bfx \bfx^{\top} | \bfz) - 
\left( \mbox{E}_{\bfz} \mbox{E}_{\bfx | \bfz} (\bfx | \bfz)   \right)
\left( \mbox{E}_{\bfz} \mbox{E}_{\bfx | \bfz} (\bfx | \bfz)^{\top}  \right). 
\]
Element $(p,p)$, $p=1,\ldots ,P$, in the diagonal of $\mbox{Var}(\bfx)$ is,
\begin{eqnarray}
\mbox{Var}(x_{.p}) &=& \mbox{E}(x_{.p}^2) - \mbox{E}(x_{.p})^2 \nonumber \\ 
&=& \sum_{x_p=1}^{M_p} x_p^2 P(x_{.p}=x_p ) - [\sum_{x_p=1}^{M_p} x_p P(x_{.p}=x_p)]^2 \nonumber \\
&=& \sum_{c=1}^{C} \psi_c [ \sum_{x_p=1}^{M_p} x_p^2 P(x_{.p}=x_p | z_i=c)  ] \nonumber \\
 & & - \{ \sum_{c=1}^{C} \psi_c [ \sum_{x_p=1}^{M_p} x_p P(x_{.p}=x_p | z_i=c)  ] \}^2 .
\end{eqnarray}
Element $(p,q)$, $p \neq q$, $p,q=1,\ldots ,P$, in the off-diagonal of $\mbox{Var}(\bfx)$ is, 
\begin{eqnarray}
\mbox{Cov}(x_{.p},x_{.q}) &=& \mbox{E}(x_{.p} \times x_{.q}) - \mbox{E}(x_{.p}) \times \mbox{E}(x_{.q}) \nonumber \\  
&=& \sum_{x_p=1}^{M_p}\sum_{x_q=1}^{M_q} x_p \times x_q P(x_{.p}=x_p, x_{.q}=x_q) \nonumber \\
 & & - \sum_{x_p=1}^{M_p} x_p P(x_{.p}=x_p) \times \sum_{x_q=1}^{M_q} x_q P(x_{.q}=x_q) \nonumber \\ 
&=& \{ \sum_{c=1}^{C} P(z_i=c) \sum_{x_p=1}^{M_p} \sum_{x_q=1}^{M_q} x_p x_q P(x_{.p}=x_p,x_{.q}=x_q | z_i=c) \} \nonumber \\ 
& & - \{ \sum_{c=1}^{C} P(z_i=c) [ \sum_{x_p=1}^{M_p} x_p P(x_{.p}=x_p | z_i=c)  ] \}  \nonumber \\ 
& & \times \{ \sum_{c=1}^{C} P(z_i=c) [ \sum_{x_q=1}^{M_q} x_q P(x_{.q}=x_q | z_i=c)  ] \}. \nonumber
\end{eqnarray}

As $x_{.p}$ and $x_{.q}$ are independent given $\bfz$,  
\begin{eqnarray}
\mbox{Cov}(x_{.p},x_{.q})&=&\sum_{c=1}^{C} P(z_i=c) [ \sum_{x_p=1}^{M_p} x_p P(x_{.p}=x_p | z_i=c)] 
[ \sum_{x_q=1}^{M_q} x_q P(x_{.q}=x_q | z_i=c)]  \nonumber \\
& & - 
\{ \sum_{c=1}^{C} P(z_i=c) [ \sum_{x_p=1}^{M_p} x_p P(x_{.p}=x_p | z_i=c)  ] \} \nonumber \\ 
& & \times 
\{ \sum_{c=1}^{C} P(z_i=c) [ \sum_{x_q=1}^{M_q} x_q P(x_{.q}=x_q | z_i=c)  ] \} \nonumber \\
&=& \sum_{c=1}^{C} \psi_c [ \sum_{x_p=1}^{M_p} x_p P(x_{.p}=x_p | z_i=c)] 
[ \sum_{x_q=1}^{M_q} x_q P(x_{.q}=x_q | z_i=c)]  \nonumber \\
& & - 
\{ \sum_{c=1}^{C} \psi_c [ \sum_{x_p=1}^{M_p} x_p P(x_{.p}=x_p | z_i=c)  ] \} \nonumber \\
& & \times 
\{ \sum_{c=1}^{C} \psi_c [ \sum_{x_q=1}^{M_q} x_q P(x_{.q}=x_q | z_i=c)  ] \} . \nonumber 
\end{eqnarray}

Denote by $f_{p,c}$ the expected value for $x_{.p}$ in cluster $c$, i.e. $f_{p,c}=\mbox{E}(x_{.p} | z_{i}=c)=\sum_{x_p=1}^{M_p} x_p P(x_{.p}=x_p | z_i=c)$. Then, for $p\neq q$, 
\begin{eqnarray}
&\mbox{Cov} (x_{.p},x_{.q}) = \sum_{c=1}^{C} \psi_c f_{p,c} f_{q,c} - \left( \sum_{c=1}^{C} \psi_c f_{p,c} \right) \left( \sum_{c=1}^{C} \psi_c f_{q,c} \right) .
\end{eqnarray}

{\bf Example 1:} Consider $C=4$, and assume that $f_{p,1}=f_{q,1}=f_{p,3}=f_{q,3}$ and 
$f_{p,2}=f_{q,2}=f_{p,4}=f_{q,4}$. Then, for $p \neq q$, it follows from (2) that,
\[
\mbox{Cov}(x_{.p},x_{.q}) = (\psi_1 + \psi_3) (\psi_2+\psi_4) (f_{p,1}-f_{p,2})^2 >0 .
\]

In the Supplementary Material, Section S2, we present an extended version of Example 1, as well as an additional example on inferences after utilizing equation (2). However, the larger the number of clusters, the less helpful (2) becomes for understanding the effect of the clustering on the marginal covariance structure of the variables. More helpful is the following Proposition. 

{\bf Proposition 1:} Assume that $x_{.p}$ and $x_{.q}$ are interval variables. Under the condition that $\psi_{1}=\psi_{2}=\ldots =\psi_{C}=\psi$, for $p \neq q$, $p,q=1,\ldots ,P$,
\begin{eqnarray}
& & \mbox{Cov}(x_{.p},x_{.q})  = \sum_{ \{c_1,c_2=1,\ldots ,C, c_1<c_2 \}} \psi^2 (f_{p,c_1} - f_{p,c_2})
(f_{q,c_1} - f_{q,c_2}).
\end{eqnarray}

{\it Proof:} The proof is given in the Appendix.

Equation (3), although restricted to $\psi_{c}=\psi$, $c=1, \dots ,C$, is more helpful for examining the effect of the clustering on the covariance structure of the variables. 
For any number of clusters, if, for all  $c_1<c_2$, the sign of $(f_{p,c_1} - f_{p,c_2})$ is the same as the sign of $(f_{q,c_1} - f_{q,c_2})$, the correlation between $x_{.p}$ and $x_{.q}$ is positive. If, for all $c_1<c_2$, the sign of $(f_{p,c_1} - f_{p,c_2})$ is different to the sign of $(f_{q,c_1} - f_{q,c_2})$, the correlation between $x_{.p}$ and $x_{.q}$ is negative. The correlation is zero if, for every term in $\mbox{Cov}(x_{.p},x_{.q})$, as given by (3), either $f_{p,c_1} = f_{p,c_2}$, or $f_{q,c_1} = f_{q,c_2}$.

\section{An algorithm for a predetermined subject clustering with partial control of the variables' dependence structure}

Let $H_p$, $p=1,\ldots ,P$, denote a probability vector that contains $M_{p}$ probabilities that add up to one. Let also $L_p$, $p=1,\ldots ,P$, denote a different probability vector that contains $M_{p}$ probabilities that add up to one. 
According to the algorithm proposed in this section, markedly different vectors $H_p$ and $L_p$ will create distinct profiles for subjects in different clusters. The algorithm generates $k=2^{C/2-1}$ groups of associated variables, where the dependence between variables within a group is stronger compared to the dependence between variables in different groups. Henceforth, we refer to those groups of variables as {\it homogenous}.  The derived dependence structure is shown empirically in Example 2, where we present sample correlations assuming interval variables. The algorithm allows for homogenous groups of different size.  
In Section 3.2, we derive a theoretical result for interval variables that allows to pre-specify within-homogenous-group covariances or correlations. In turn, this specification determines what $H_p$ and $L_p$ should be. We determine $C$ to be even, as this generates a clearly defined dependence structure. This is shown in Examples S3 and S4 in Section S3 of the Supplementary Material, where a variation of the proposed algorithm is considered that allows for odd $C$.  The variables are positively correlated within each homogenous group of variables. 

\subsection{The proposed algorithm}

The proposed algorithm is shown below. Explanatory comments are added in brackets. 

\begin{itemize}
\item[(1)] Define the number $k$ of homogenous groups of variables, where $k$ is a power of 2. Solving $k=2^{C/2-1}$ provides the even number of clusters, $C=2*[ln(k)/ln(2) +1]$.

\item[(2)] Define the number of variables $l_v$ in each homogenous group $v$, $v=1,\ldots ,k$.  

\item[(3)] Define the number of subjects, $n_1=\ldots =n_C$, within each cluster. 

\item[(4)] For each variable $x_{.p}$, consider two sets of probabilities, \\ $H_p=\{\phi^{H}_{p}(1),\ldots ,\phi^{H}_{p}(M_{p})\}$, and, $L_p=\{\phi^{L}_{p}(1),\ldots ,\phi^{L}_{p}(M_{p})\}$, so that, \\ $\sum_{m=1}^{M_{p}} \phi^{H}_{p}(m)=1$, and, 
$\sum_{m=1}^{M_{p}} \phi^{L}_{p}(m)=1$. The two sets could be distinct so that the first elements of $H_p$ are considerably higher than subsequent elements, whilst the first elements of $L_p$ are considerably lower. 

\item[(5)] For odd $c$, define the profile of cluster $c$ so that:
\begin{itemize}
\item the first $l_1+\ldots +l_{k/(2^{c/2-0.5})}$ variables are simulated in accordance with $\{ L_1,\ldots ,L_P \}$
\item the next $l_{k/(2^{c/2-0.5})+1}+\ldots +l_{k/(2^{c/2-0.5})+k/(2^{c/2-0.5})}$ variables in accordance with $\{ H_1,\ldots ,H_P \}$ 
\item the next $l_{k/(2^{c/2-0.5})+k/(2^{c/2-0.5})+1}+\ldots +l_{k/(2^{c/2-0.5})+k/(2^{c/2-0.5})+k/(2^{c/2-0.5})}$ variables in accordance with $\{ L_1,\ldots ,L_P \}$
\item and so on and so forth. 
\end{itemize}

\item[(6)] For even $c$, define the profile of cluster $c$ so that:
\begin{itemize}
\item the first $l_1+\ldots +l_{k/(2^{c/2-1})}$ variables are simulated in accordance with $\{ H_1,\ldots ,H_P \}$ 
\item the next $l_{k/(2^{c/2-1})+1}+\ldots +l_{k/(2^{c/2-1})+k/(2^{c/2-1})}$ variables in accordance with $\{ L_1,\ldots ,L_P \}$ 
\item the next $l_{k/(2^{c/2-1})+k/(2^{c/2-1})+1}+\ldots +l_{k/(2^{c/2-1})+k/(2^{c/2-1})+k/(2^{c/2-1})}$ variables in accordance with $\{ H_1,\ldots ,H_P \}$ 
\item and so on and so forth.
\end{itemize}

\item[] \{ When $l_1=\ldots =l_k$, the two steps above simplify as follows: for odd $c$, define the profile of cluster $c$ so that the first $P/(2^{c/2-0.5})$ variables are simulated in accordance with $\{ L_1,\ldots ,L_P \}$, the next $P/(2^{c/2-0.5})$ variables considering $\{ H_1,\ldots ,H_P \}$, and so on and so forth. For even $c$, the first $P/(2^{c/2-1})$ variables are simulated considering $\{ L_1,\ldots ,L_P \}$, the next $P/(2^{c/2-1})$ variables in accordance with $\{ H_1,\ldots ,H_P \}$, and so on and so forth. \}

\item[(7)] If required, to generate observations from variables $x_{.q}$, $q>P$, that do not contribute to the clustering, consider $A_q=\{\phi^{A}_{q}(1),\ldots ,\phi^{A}_{q}(M_{q})\}$, distinct from $H_p$ and $L_p$. For all subjects, generate observations from $A_q$ irrespectively of cluster allocation. 

\end{itemize}

Proposition 2, elucidates the derived covariance structure for interval variables. 

{\bf Proposition 2:} For the algorithm proposed in Section 3.1, and for $H_1=\ldots =H_P=H$, and, $L_1=\ldots =L_P=L$, the covariance between variables within a homogenous group is the same for all groups, and is higher than any covariance between variables that belong to different groups. 

Proof: Without any loss of generality, assume that all variables contribute to the clustering. Each of the $2^{C/2 - 1}$ homogenous groups contains $l=P/(2^{C/2 - 1})$ adjoined variables with the same cluster profile characterized by H or L.  
For the variables within a homogenous group, 
the differences $(f_{p,c_1} - f_{p,c_2})$ and $(f_{q,c_1} - f_{q,c_2})$ always carry the same sign, for any $c_1$ and $c_2$. This is not true for variables in different groups. This translates to within-group covariances $\mbox{Cov}(x_{.p},x_{.q})$ that are always positive and larger than between-group covariances, as the algorithm determines balanced sized clusters and Proposition 2 holds. 

{\bf Example 2:} Assume 6 clusters ($C=6$), 12 variables ($P=12$) and $l_1=2$, $l_2=2$, $l_3=5$, and $l_4=3$. Note that $k=4$. Consider $600$ subjects. Observations were simulated using $H=\{0.9025,0.0950,0.025\}$ and $L=\{0.0625,0.3750,0.5625\}$. In Table 1, we present the cluster profiles created by the proposed algorithm. Specifically,
\begin{itemize}
\item Cluster 1: $c=1$ and $k/(2^{c/2-0.5})=4/1=4$. Then, $l_1+\dots +l_4=12$ and, according to step [5], observations from $\{1,2,3 \}$ are simulated using the probabilities in vector $L$ for all variables. 
\item Cluster 2: $c=2$ and $k/(2^{c/2-1})=4/1=4$. Then, $l_1+\dots +l_4=12$ and, according to step [6], observations from $\{1,2,3 \}$ are simulated using the probabilities in vector $H$ for all variables.
\item Cluster 3: $c=3$ and $k/(2^{c/2-0.5})=4/2=2$. According to step [5], observations for the first $l_1+l_2=4$ variables are simulated using the probabilities in vector $L$, whilst observations for the remaining $l_3+l_4=8$ variables are simulated according to $H$. 
\item Cluster 4: $c=4$ and $k/(2^{c/2-1})=4/2=4$. According to step [6], observations for the first $l_1+l_2=4$ variables are simulated using $H$, and for the remaining $l_3+l_4=8$ variables using $L$.
\item Cluster 5: $c=5$ and $k/(2^{c/2-0.5})=4/4=1$. According to step [5], observations for the first $l_1=2$ variables are simulated using $L$, for the next $l_2=2$ variables using $H$, for the next $l_3=5$ variables using $L$, and for the last $l_4=3$ variables using $H$.
\item Cluster 6: $c=6$ and $k/(2^{c/2-1})=4/4=1$. According to step [6], observations for the first $l_1=2$ variables are simulated using $H$, for the next $l_2=2$ variables using $L$, for the next $l_3=5$ variables using $H$, and for the last $l_4=3$ variables using $L$.
\end{itemize}

In Figure 1(a), 
 we present a heatmap of the theoretical correlations assuming interval variables, and in Figure 1(b) 
 the sample correlations. Note that blocks of negative and zero correlations are observed in the correlation matrix, due to the symmetry in the clustering structure.  The clustering of the simulated data is in accordance with the predetermined clustering; see Figure S7 in Section S4 of the Supplementary Material. (Throughout the manuscript, simulated subject profiles are clustered using the R package PReMiuM (Liverani et al. 2015), which implements Bayesian clustering with the Dirichlet process.) This is observed in subsequent examples too, as well as the examples in the Supplementary Material. 

\begin{small}
\begin{center}
\begin{table*}[!t]
{\bf Table 1:} {Cluster profiles for 12 variables ($P=12$) and 6 clusters ($C=6$) for Example 2. Observations are simulated using  probability vectors $L$ and $H$.}
\label{tab:5}
\par
\begin{tabular}{lcccccccccccc}
\hline 
  & $x_{.1}$ & $x_{.2}$ & $x_{.3}$ & $x_{.4}$ & $x_{.5}$ & $x_{.6}$ & $x_{.7}$ & $x_{.8}$ 
  & $x_{.9}$ & $x_{.10}$ & $x_{.11}$ & $x_{.12}$ \\
\hline
Cluster 1 & L & L & L & L & L & L & L & L & L & L & L & L   \\
Cluster 2 & H & H & H & H & H & H & H & H & H & H & H & H   \\
Cluster 3 & L & L & L & L & H & H & H & H & H & H & H & H   \\
Cluster 4 & H & H & H & H & L & L & L & L & L & L & L & L   \\
Cluster 5 & L & L & H & H & L & L & L & L & L & H & H & H    \\
Cluster 6 & H & H & L & L & H & H & H & H & H & L & L & L   \\
\hline 
\end{tabular}
\end{table*}
\end{center}
\end{small}

\subsection{Allowing for a predetermined covariance or correlation within each homogenous group for interval SNP-like variables}

{\bf Proposition 3:} Assume that interval variables $x_{.p}$ and $x_{.q}$ belong to the same homogenous group. For the algorithm in Section 3.1, 
\[
\mbox{Cov}(x_{.p},x_{.q}) =  0.25 \times (f_{(p,H)} - f_{(p,L)})\times (f_{(q,H)} - f_{(q,L)}), 
\]
where, $f_{(p,H)}=\sum_{x_p=1}^{M_p} x_p P(x_{.p}=x_p | H_p)$, and, $f_{(p,L)}=\sum_{x_p=1}^{M_p} x_p P(x_{.p}=x_p | L_p)$. 

{\it Proof:} See Appendix. 

In practice, one may consider the simplified scenario where variables in the same homogenous group share the same set of possible values, and $(f_{(p,H)} - f_{(p,L)}) = (f_{(q,H)} - f_{(q,L)})$. Then, given $\mbox{Cov}(x_{.p},x_{.q})$, one can set cluster specific probabilities so that, for all $x_{.p}$ in the same homogenous group,  
\begin{eqnarray}
|f_{(p,H)} - f_{(p,L)}| = \sqrt{4 \mbox{Cov}(x_{.p},x_{.q})},
\end{eqnarray}
where $|.|$ denotes absolute value. Proposition 3 and the result above can be used for the determination of marginal covariances and correlations for interval variables with any number of levels, as the proof of Proposition 3 applies generally. We now show how to utilise the results above for simulating SNP-like variables. 

{\bf \underline{Application to SNP variables, given predetermined covariances}:} \\ Single Nucleotide Polymorphisms (SNP) are  observations with 3 levels, usually denoted by $0,1$ and $2$ for `Wild type', `Heterozygous variant' and `Homozygous variant' respectively. For a SNP $x_{.p}$, due to the Hardy-Weinberg principle, (Ziegler \& K\"{o}nig, 2010),
$P(x_{.p}=0)=p_{S_p}^2$, $P(x_{.p}=1)=2 p_{S_p} (1-p_{S_p})$ and $P(x_{.p}=2)=(1-p_{S_p})^2$, where $0<p_{S_p}<1$.  Thus, $\mbox{E}(x_{.p} | z_i=c) = 2-2 p_{S_p}$, $\mbox{E}(x_{.p}^2 | z_i=c) = (1-p_{S_p}) (4-2 p_{S_p})$,  
$\mbox{Var}(x_{.p} | z_i=c) = 2 p_{S_p} (1-p_{S_p})$, and, 
$f_{(p,H)} - f_{(p,L)} = E(x_{.p} | p^{H}_{S_p}) - E(x_{.p} | p^{L}_{S_p}) = 2 (p^{H}_{S_p} - p^{L}_{S_p})$, where $p^{H}_{S_p}$ and $p^{L}_{S_p}$ are the probabilities that form the $H$ and $L$ SNP probability vectors. 
Assume that for $x_{.p}$ and $x_{.q}$ in the same homogenous group, $p^{H}_{S_p}=p^{H}_{S_q}$, and,  $p^{L}_{S_p}=p^{L}_{S_q}$, and therefore, 
$f_{(p,H)}=f_{(q,H)}$ and $f_{(p,L)}=f_{(q,L)}$. 
From (5), given a required covariance $\mbox{Cov}(x_{.p},x_{.q})$, set cluster specific probabilities for $x_{.p}$ and $x_{.q}$ so that, $2 |p^{H}_{S_p} - p^{L}_{S_p}| = \sqrt{4 \mbox{Cov}(x_{.p},x_{.q})}$. 
In practice, set $p^{H}_{S_p}=p^{H}_{S}$ suitably high and constant for all variables (say, $p^{H}_{S}\simeq 1$), and allow $p^{L}_{S_p}$ to vary in accordance with, $p^{H}_{S} - p^{L}_{S_p} = \sqrt{\mbox{Cov}(x_{.p},x_{.q})}$. 

{\bf Example 3:} Assume 6 clusters ($C=6$), 12 variables ($P=12$) that emulate SNPs, and $l_1=2$, $l_2=2$, $l_3=5$, $l_4=3$. Consider $600$ subjects, and $p^{H}_{S}=0.95$. Assume a covariance of 0.45 for the variables within homogenous groups.  In Figure 2(a), we present a heatmap of the theoretical correlation matrix for the specifications in this example, whilst sample correlations are shown in Figure 2(b). 

{\bf \underline{Application to SNP variables, given predetermined correlations}:} \\ 
From Section 2.3, equation (1), and for $\psi_1=\ldots =\psi_C=\psi$,   
\[
\mbox{Var}(x_{.p}) = \psi \sum_{c=1}^{C} [ \mbox{E}(x_{.p}^2 | z_i=c)  ] - 
\psi^2 \{ \sum_{c=1}^{C} [  \mbox{E}(x_{.p} | z_i=c) ] \}^2 . 
\]
For even $C$, for half of the clusters, $\mbox{E}(x_{.p}^2 | z_i=c)=(1-p^{H}_{S_p}) (4-2 p^{H}_{S_p})$ and 
$\mbox{E}(x_{.p} | z_i=c)=2-2 p^{H}_{S_p}$. For the remaining clusters, $\mbox{E}(x_{.p}^2 | z_i=c)=(1-p^{L}_{S_p}) (4-2 p^{L}_{S_p})$, and $\mbox{E}(x_{.p} | z_i=c)=2-2 p^{L}_{S_p}$. Therefore, $\mbox{Var}(x_{.p})$ is given by, 
\begin{eqnarray}
&\psi& [\frac{C}{2} (1-p^{H}_{S_p}) (4-2 p^{H}_{S_p}) + \frac{C}{2} (1-p^{L}_{S_p}) (4-2 p^{L}_{S_p}) ] - \psi^2 [\frac{C}{2} (2-2p^{H}_{S_p}) + \frac{C}{2} (2-2p^{L}_{S_p})]^2 \nonumber \\
&=& \psi C (1-p^{H}_{S_p}) (2- p^{H}_{S_p}) + \psi C (1-p^{L}_{S_p}) (2- p^{L}_{S_p}) - 
\psi^2 C^2 (2-p^{H}_{S_p}-p^{L}_{S_p})^2. \nonumber 
\end{eqnarray}
Then, as $\psi=1/C$,
\begin{eqnarray}
& & \mbox{Cov}(x_{.p},x_{.q}) \nonumber \\ 
&=& \mbox{Cor}(x_{.p},x_{.q}) [
 (1-p^{H}_{S_p}) (2- p^{H}_{S_p}) + (1-p^{L}_{S_p}) (2- p^{L}_{S_p}) - (2-p^{H}_{S_p}-p^{L}_{S_p})^2 ].  \nonumber
\end{eqnarray}
As we demonstrated earlier in this Section, for a given $\mbox{Cov}(x_{.p},x_{.q})$, 
$p^{H}_{S} - p^{L}_{S_p} = \sqrt{\mbox{Cov}(x_{.p},x_{.q})}$. 
Thus, to allow for different predetermined correlations within each homogenous group of variables, one should set set $p^{H}_{S_p}=p^{H}_{S}$ suitably high (say, close to 1), and let $p^{L}_{S_p}$ vary so that, 
\begin{eqnarray}
& & p^{H}_{S} - p^{L}_{S_p} \nonumber \\ 
&=&  \sqrt{ \mbox{Cor}(x_{.p},x_{.q})}  \times  
\sqrt{[ (1-p^{H}_{S}) (2- p^{H}_{S}) + (1-p^{L}_{S_p}) (2- p^{L}_{S_p}) - 
(2-p^{H}_{S}-p^{L}_{S_p})^2 ] }. \nonumber
\end{eqnarray}
Note that the chosen correlation is restricted so that, for $p^{L}_{S_p} \rightarrow 0$, the maximum possible correlation is $p^{H}_{S}$. The restriction is negligible for $(p^{H}_{S}) \simeq 1$.

{\bf Example 4:} Assume 8 clusters ($C=8$), 16 variables ($P=16$) that emulate SNPs, and $l_v=2$, $v=1,\ldots ,8$. Consider $800$ subjects, with observations simulated using $p^{H}_{S}=0.95$, for predetermined correlations within the 8 homogenous groups given by $(0.4,0.5,0.6,0.7,0.8,0.6,0.7,0.4)$.  In Figure 3(a), we present a heatmap of the theoretical correlation matrix, for the specifications in this example. In Figure 3(b) we present the sample correlation matrix for the simulated observations.

\section{Genetic profiles defined by correlated SNPs - A GWA study}

Data from a GWA study of lung cancer (Hung et al. 2008) are utilized. Genotyping was performed with the Illumina Sentrix HumanHap300 BeadChip, including 317,139 SNPs of subjects from the International Agency for Research on Cancer (IARC) lung cancer study. The top 200 SNPs, ranked by their 
p-value for association with lung cancer (adjusted for age, sex, and country) were selected. The correlation (Linkage Disequilibrium) structure is shown in Figure 4(a). We observe 27 groups of SNPs, where SNPs are correlated within each group and uncorrelated between groups. Correlations are overwhelmingly positive. Table 2, shows the average sample correlation within each of the 27 groups, for the 89 SNPs that are correlated with at least one other polymorphism. 

The algorithm in Section 3.1 is used to generate a predetermined clustering structure for 6000 subjects, using simulated observations from 200 SNPs, whilst the specified homogenous groups resemble those in the real data set. 
For 12 predetermined clusters, we consider 32 homogenous groups of SNPs. For the first 27 groups, we specify within-group correlations that match the within-group correlations in the real data set. For the last five groups, created to satisfy the requirements of the proposed algorithm, we determine a very small within-group correlation of 0.01. This is because each one of the 10 SNPs in the last 5 groups corresponds to a SNP in the real data that is not correlated with any other SNP. The clustering structure in the simulated data is exactly as pre-determined, with 12 clusters containing 500 subjects each (Supplementary Material, Section S5, Figure S8). Within-group sample correlations for the simulated data are shown in Table 2. The simulated dataset replicates almost exactly the real within-group correlations. Such control is a considerable improvement compared to the standard algorithm described in Section 2. 

The Linkage Disequilibrium structure within the simulated data set is shown in Figure 4(b). Due to the symmetry in the clustering algorithm, we observe a notable simulated correlation structure between homogenous groups, not observed in the real dataset. Figure S9, in the Supplementary Material, Section S5, shows this more clearly, as the focus is on the first 99 SNPs, ignoring the last 101 uncorrelated Polymorphisms. In the next section, we discuss in more detail the issue of controlling between-group correlations independently of within-group correlations. 

\begin{center}
\begin{table*}[!t]
{\bf Table 2:} {Average within-group sample correlations for the 89 correlated SNPs from Hung et al. (2008), and for the simulated data. In parentheses the number of SNPs in each group.}
\label{tab:1}
\par
\begin{tabular}{ccccccccccc}
\hline
Group     & 1 (2) & 2 (2) & 3 (3) & 4 (2) & 5 (2) & 6 (3) & 7 (2) & 8 (2) & 9 (3) & 10 (2) \\
Corr (LD) - Real & 0.68 & 0.96 & 0.62 & 0.91 & 0.96 & 0.93 & 0.90 & 0.98 & 0.91 & 0.98 \\
Corr (LD) - Sim & 0.69 & 0.96 & 0.63 & 0.90 & 0.96 & 0.93 & 0.90 & 0.98 & 0.91 & 0.98 \\
\hline 
Group     & 11 (2) & 12 (2) & 13 (21) & 14 (5) & 15 (2) & 16 (2) & 17 (3) & 18 (2) & 19 (3) & 20 (4) \\
Corr (LD) - Real  & 0.96 & $0.98^{*}$ & 0.59 & 0.94 & 0.32 & 0.92 & 0.41 & 0.96 & 0.63 & 0.66 \\
Corr (LD) - Sim   & 0.96 & $0.98$ & 0.59 & 0.94 & 0.31 & 0.92 & 0.41 & 0.96 & 0.64 & 0.66 \\
\hline 
Group     & 21 (2) & 22 (7) & 23 (2) & 24 (3) & 25 (2) & 26 (2) & 27 (2) &  &  &  \\
Corr (LD) - Real  & 0.96 & $0.60^{+}$ & 0.42 & 0.90 & 0.43 & 0.56 & 0.74 & &  & \\
Corr (LD) - Sim   & 0.96 & $0.60$ & 0.43 & 0.90 & 0.41 & 0.55 & 0.74 & & &  \\
\hline 
\end{tabular}

* Actual correlation is 0.99. 0.98 used to avoid numerical instability \\
+ Actual average correlation is 0.15. 0.6 used, after excluding negative within-group correlations 
\end{table*}
\end{center}

\section{Discussion}

Our work concerns interval variables. Empirical evidence shows that the proposed algorithm generates a similar dependence structure for ordinal observations. The dependence structure considering nominal data differs, as negative associations are not present. Nevertheless, we observed in various examples that the overall structure of positive associations was quite similar between ordinal/interval and nominal variables, albeit weaker for the latter. All empirical evidence suggests that the manner in which control is effected over within-group correlations is also relevant to nominal and ordinal variables, in terms of the comparative magnitude of within-group associations. See Section S6 in the Supplementary Material for more details.  

The algorithm described in Section 3 allows for a predetermined clustering structure for the subjects, whilst assuming a specific stratified exchangeable structure for the marginal correlations of the variables. This assumption is obviously restrictive, as other marginal dependence structures may be observed. However, the algorithm allows to specify the size of each one of the homogenous groups, and the value of each one of the within group correlations. This makes it flexible enough to define a large variety of clustering and marginal dependence structures. Specifically, the user is free to define either the number of clusters $C$, or the number of homogenous groups of variables $k$ as a power of $2$. This appears to be inflexible, as one quantity then appears to define the other through $k=2^{C/2-1}$. However, freely choosing the number of clusters only places an upper bound on the number of homogenous groups of variables. This is because correlations within some of the homogenous groups can be effectively zero. 
In addition, it is not essential that $C$ is set to be even; see Supplementary material Section S3. We consider even $C$ as this creates a more distinct clustering structure.  The algorithm's flexibility is further maintained as the investigator is free to choose the number of subjects, and the number of variables within each homogenous group.  

The two sets of probabilities $H$ and $L$ are sufficient for defining distinct cluster profiles, due to the proposed profile structure illustrated in Table 1. Using more than two sets of probabilities would add unnecessary complexity to the algorithm. 

Cluster sizes are assumed equal to derive theoretical results on the algorithm's properties, but this is not essential for the implementation of the algorithm. Empirical evidence has shown that under unequal cluster sizes, the dependence structure between the variables created by the proposed algorithm is similar to the one derived theoretically. For one such example see the Supplementary material, Section S7. 

It is well known that the correlation of interval variables is restricted in accordance with marginal probabilities. The most straightforward and trivial example is binary variables. For instance, for marginal probabilities $P(x_{.1}=1)=0.2$ and $P(x_{.2}=1)=0.8$, the maximum possible correlation is $0.25$, attained for $P(x_{.1}=1,x_{.2}=1)=0.2$. We saw in Section 3.2 that the pre-defined correlation is indeed constrained by the choice of $p_{S}^{H}$, but the restriction is negligible for $(p^{H}_{S}) \simeq 1$.

The proposed algorithm effects control over within-group correlations. Between-group correlations are present as a direct consequence of the symmetry in the clustering structure. Determination of between-group correlations independently of the within-group structure, in tandem with the predetermined clustering, is not straightforward. Equation (3) offers a direct link between the covariances $\mbox{Cov}(x_{.p},x_{.q})$, and the variable profiles in each cluster, through $f_{p,c}$, $p=1,\ldots ,P$, $c=1,\ldots ,C$. $P$ variables imply $P(P-1)/2$ covariances, under the constraint that they form a positive definite matrix. The number of different $(f_{p,c_1} - f_{p,c_2})$ quantities is ${C \choose 2} P$. It is straightforward to deduce that the number of unconstrained $(f_{p,c_1} - f_{p,c_2})$ quantities is $P(C-1)$. For predetermined covariances, (3) generates a non-linear system of $P(P-1)/2$ equations, with $P(C-1)$ unknowns. Solving such a system could, in principle, allow to set between-group correlations independently of within-group associations. However, this approach is not reliable. Numerical solutions for simple examples are not available, with no solution or an infinite number of solutions reported by the symbolic computation software MAPLE. For a specific example, consider $P=5$, $C=2$, $\psi=0.5$, and a covariance structure so that,  
$\mbox{Cov}(x_{.1},x_{.2})=\mbox{Cov}(x_{.1},x_{.3})=\mbox{Cov}(x_{.2},x_{.3})=0.5$, 
$\mbox{Cov}(x_{.4},x_{.5})=0.6$ and 
$\mbox{Cov}(x_{.1},x_{.4})=\mbox{Cov}(x_{.1},x_{.5})=\mbox{Cov}(x_{.2},x_{.4})=  
\mbox{Cov}(x_{.2},x_{.5})=\mbox{Cov}(x_{.3},x_{.4})=\mbox{Cov}(x_{.3},x_{.5})=0.2$. This  
provides a system of equations with no solution according to MAPLE. A specification where $P=4$, $C=4$, $\psi=0.25$, and covariances zero except of $\mbox{Cov}(x_{.1},x_{.2})=\mbox{Cov}(x_{.3},x_{.4})=0.49$, generates a system with an infinite number of solutions. Solving the system of equations produced by (3) can be problematic even when the system includes equal numbers of equations and unknowns. For instance, $P=5$ and $C=3$ creates a system of equations with a Jacobian equal to zero and an infinite number of solutions.  This suggests that a generally applicable algorithm, such as the one proposed in Section 3, is a suitably pragmatic approach for achieving control over the marginal dependence of the variables.


\vspace{0.2cm}
\noindent {\bf Data availability}
\vspace{0.1cm}

The data and code are available from the author upon request.

\vspace{0.2cm}
\noindent {\bf Acknowledgment}
\vspace{0.1cm}

\noindent 
We would like to thank Professor Paolo Vineis and Dr Paul Brennan for providing the data used in Section 5.

{\bf APPENDIX: Proof of Proposition 1:} \\
From (2),
\begin{eqnarray}
&\mbox{Cov}&(x_{.p},x_{.q}) = \sum_{c=1}^{C} \psi f_{p,c} f_{q,c} 
 - \left( \sum_{c=1}^{C} \psi f_{p,c} \right) \left( \sum_{c=1}^{C} \psi f_{q,c} \right) \nonumber \\ 
&=& \sum_{c=1}^{C} \psi f_{p,c} f_{q,c} - \sum_{c=1}^{C} \psi^2 f_{p,c} f_{q,c} \nonumber \\
 & & - \sum_{c_1<c_2, c_2=2,\ldots ,C} \psi^2 f_{p,c_1} f_{q,c_2} 
- \sum_{c_1<c_2, c_2=2,\ldots ,C} \psi^2 f_{p,c_2} f_{q,c_1}  \nonumber \\ 
&=& \sum_{c=1}^{C} (\psi - \psi^2) f_{p,c} f_{q,c} \nonumber \\ 
& & - \sum_{c_1<c_2, c_2=2,\ldots ,C} \psi^2 f_{p,c_1} f_{q,c_2} - \sum_{c_1<c_2, c_2=2,\ldots ,C} \psi^2 f_{p,c_2} f_{q,c_1}.
\end{eqnarray}
Now, 
\begin{eqnarray}
& &\sum_{ c_1<c_2, c_2=2,\ldots ,C }  (\psi f_{p,c_1} - \psi f_{p,c_2})
(\psi f_{q,c_1} - \psi f_{q,c_2}) \nonumber \\
&=& \sum_{c_1<c_2, c_2=2,\ldots ,C} [ \psi^2 f_{p,c_1} f_{q,c_1} + \psi^2 f_{p,c_2} f_{q,c_2} 
 - \psi^2 f_{p,c_1} f_{q,c_2} - \psi^2 f_{p,c_2} f_{q,c_1}] \nonumber \\
&=& \sum_{c=1}^{C} (C-1) \psi^2 f_{p,c} f_{q,c} \nonumber \\ 
& & - \sum_{c_1<c_2, c_2=2,\ldots ,C}  \psi^2 f_{p,c_1} f_{q,c_2} 
- \sum_{c_1<c_2, c_2=2,\ldots ,C} \psi^2 f_{p,c_2} f_{q,c_1}. 
\end{eqnarray}
To complete the proof we show that (6)=(7), i.e. that, 
\[
\sum_{c=1}^{C} (\psi - \psi^2) f_{p,c} f_{q,c} = \sum_{c=1}^{C} (C-1) \psi^2 f_{p,c} f_{q,c}. 
\]
To show this, notice that, 
\[
\psi - \psi^2=  \psi (1- \psi)  =\psi (C-1) \psi = (C-1) \psi^2 
\]
and the proof of Proposition 1 is complete.

{\bf APPENDIX: Proof of Proposition 3:} \\
For the algorithm in Section 3.1, $\psi_{1}=\psi_{2}=\ldots =\psi_{C}=\psi=1/C$, so that Propositions 1 and 2 hold. From Proposition 1, 
\begin{eqnarray}
\mbox{Cov}(x_{.p},x_{.q}) = \sum_{ \{c_1,c_2=1,\ldots ,C, c_1<c_2 \}} \psi^2 (f_{p,c_1} - f_{p,c_2})
(f_{q,c_1} - f_{q,c_2}). \nonumber 
\end{eqnarray}
The number of terms in the right hand side sum is ${C \choose 2}$. For the algorithm in Section 3.1, and for all $p=1,\ldots ,P$, all non-zero terms $(f_{p,c_1} - f_{p,c_2})$ are equal in absolute value. We denote this absolute value by $|f_{(p,H)} - f_{(p,L)}|$, where, $f_{(p,H)}=\sum_{x_p=1}^{M_p} x_p P(x_{.p}=x_p | H_p)$, and, $f_{(p,L)}=\sum_{x_p=1}^{M_p} x_p P(x_{.p}=x_p | L_p)$. The number of non-zero terms in either $f_{(p,H)}$ or $f_{(p,L)}$ is,
\[
\sum_{i=1}^{C/2} i + \sum_{i=1}^{(C-2)/2} i.
\]
[This can be deduced by first picking two variables from the same homogenous group. Then consider the Table that shows the cluster profiles, as in Table 1. Start from the top row of the Table and count the non-zero terms moving down the Table rows. Repeat, starting from the second row, counting the non-zero terms down the rows and so on and so forth.] 
\begin{eqnarray}
\sum_{i=1}^{C/2} i + \sum_{i=1}^{(C-2)/2} i &=& \frac{(\frac{C}{2}+1) \frac{C}{2}}{2} + 
 \frac{(\frac{C-2}{2}+1) \frac{C-2}{2}}{2}  \nonumber \\
&=& \frac{(C+2) C}{2\times 4} + \frac{(C-2+2)C-2}{2\times 4} \nonumber \\ 
&=& \frac{(C+2)C+C(C-2)}{2 \times 4} \nonumber \\
&=& \frac{C (2 C)}{8} = \frac{C^2}{4}. \nonumber  
\end{eqnarray}

For variables $x_{.p}$ and $x_{.q}$ in the same homogenous group, 
$(f_{p,c_1} - f_{p,c_2})$ and $(f_{q,c_1} - f_{q,c_2})$ always carry the same sign. Therefore, 
\begin{eqnarray}
(f_{p,c_1} - f_{p,c_2}) \times (f_{q,c_1} - f_{q,c_2}) 
 = |f_{(p,H)} - f_{(p,L)}| \times |f_{(q,H)} - f_{(q,L)}|. \nonumber 
\end{eqnarray}
Thus, we can write, 
\begin{eqnarray}
\mbox{Cov}(x_{.p},x_{.q}) &=&  \psi^2 \frac{C^2}{4} (f_{(p,H)} - f_{(p,L)})  (f_{(q,H)} - f_{(q,L)}) \nonumber \\
&=&  0.25 (f_{(p,H)} - f_{(p,L)}) \times (f_{(q,H)} - f_{(q,L)}). \nonumber 
\end{eqnarray}
This completes the proof of Proposition 3. 

\parindent 0mm \parskip 0cm
\vspace{0.2cm}
\noindent {\bf References}
\vspace{0.1cm}

\begin{list}{}
{\setlength{\itemsep}{0cm}
\setlength{\parsep}{0cm}
\setlength{\leftmargin}{0.5cm}
\setlength{\labelwidth}{0.5cm}
\setlength{\itemindent}{-0.5cm}
}


\item Allman, E.S., Matias, C., Rhodes, J.A. (2009). Identifiability of parameters in latent structure models with many observed variables. {\it Annals of Statistics}, 37, 3099-3132.

\item Bhattacharya, A., Dunson, D.B. (2012). Simplex Factor Models for Multivariate Unordered
Categorical Data. {\it Journal of the American Statistical Association}, 107, 362-377.


\item Bingham, S., Riboli, E. (2004). Diet and cancer - the European prospective Investigation into cancer and nutrition. {\it Nature Reviews Cancer}, 4, 206-215.

\item Celeux, G., Govaert, G. (2016). Latent class models for categorical data. In: Hennig C, Meila M, Murtagh F, Rocci R (ed) Handbook of cluster analysis. Handbooks of modern statistical methods. Chapman \& Hall/CRC Press, pp 173-194

\item Dunson, D.B., Xing, C. (2009). Nonparametric Bayes modeling of multivariate categorical data. {\it Journal of the American Statistical Association}, 104, 1042-1051. 

\item Fr{\"u}hwirth-Schnatter. S. (2016). {\it  Finite Mixture and Markov Switching Models.} Springer

\item Gr\"un, B., Malsiner-Walli, G. (2022) Bayesian finite mixture
models. In N. Balakrishnan, Theodore Colton, Brian Everitt, Walter Piegorsch, Fabrizio
Ruggeri, and Jef L. Teugels, editors, Wiley StatsRef: Statistics Reference Online doi:10.
1002/9781118445112.stat08373.

\item Hennig, C. (2015). What are the true clusters? {\it Pattern Recognition Letters} 64, 53-62

\item Hennig, C., Meila, M., Murtagh, F., Rocci, R. (ed) (2015). {\it Handbook of cluster analysis.} Chapman \& Hall/CRC Press 

\item Hung, R.J., McKay, J.D., Gaborieau, V., Boffetta, P., Hashibe, M., Zaridze, D. et al. (2008). A susceptibility locus for lung cancer maps to nicotinic acetylcholine receptor subunit genes on 15q25. {\it Nature}, 452, 633-637. 

\item Ioannidis, J.P.A., Thomas, G., Daly, M.J. (2009). Validating, augmenting and refining genome-wide association signals. {\it Nature Reviews}, 10, 318-329. 

\item Jasra, A., Holmes, C.C., Stephens, D.A. (2005). Markov Chain Monte Carlo Methods and the Label Switching Problem in Bayesian Mixture Modeling. {\it Statistical Science}, 20, 50-67. 

\item Jing, W., Papathomas, M. Liverani, S. (2024). Variance Matrix Priors for Dirichlet Process Mixture Models With Gaussian
Kernels. {\it International Statistical Review},  doi: 10.1111/insr.12595

\item Kirk, P., Pagani, F. Richardson, S. (2023). Bayesian outcome-guided multi-view mixture models with applications in molecular precision medicine. 	arXiv:2303.00318

\item Liverani, S., Hastie, D.I., Azizi, L., Papathomas, M., Richardson, S. (2015). PReMiuM: An R package for profile regression mixture models using Dirichlet processes. {\it Journal of Statistical Software}, 64, 1-30. 

\item Lorenz, D.J., Datta, S., Harkema, S.J. (2011). Marginal association measures for clustered data. {\it Statistics in Medicine}, 30, 3181-3191.

\item Marbac, M., Biernacki, C., Vandewalle, V. (2014). Model-based clustering for conditionally correlated categorical data.  arXiv:1401.5684v2 


\item Molitor, J., Papathomas, M., Jerrett, M., Richardson, S. (2010). Bayesian profile regression with an application to the National Survey of Children's Health. {\it Biostatistics}, 11, 484-98. 

\item M\"{u}ller, P., Quintana, F., Rosner, G.L. (2012). A Product Partition Model with Regression on
Covariates. {\it Journal of Computational and Graphical Statistics}, 20, 260-278. 

\item Oberski, D.L. (2016). Beyond the number of classes: separating substantive from non-substantive dependence in latent class analysis. {\it Advances in Data Analysis and Classification}, 10, 171-182. 

\item Papathomas, M., Molitor, J., Riboli, E., Richardson, S., Vineis, P. (2011). Examining the joint effect of multiple risk factors using exposure risk profiles: lung cancer in non-smokers. {\it Environmental Health Perspectives}, 119, 84-91. 


\item Papathomas, M., Richardson, S. (2016). Exploring dependence between categorical variables: benefits and limitations of using variable selection within Bayesian clustering in relation to log-linear modelling with interaction terms. {\it Journal of Statistical Planning and Inference}, 173, 47-63. 

\item Paulou, M., Seaman, S.R., Copas, A.J. (2013). An examination of a method for marginal inference when the cluster size is informative. {\it Statistica Sinica}, 23, 791-808.



\item Wang, A., Sabo, R.T. (2015). Simulating clustered and dependent binary variables. {\it Biostatistics}, Theory and Methods 2, 1-5. 


\item Yau, C., Holmes, C. (2011). Hierarchical Bayesian nonparametric mixture models for clustering with variable relevance determination. {\it Bayesian Analysis}, 6, 329-352.  

\item Zhou, J., Bhattacharya, A., Herring, A.H., Dunson, D.B. (2015). Bayesian factorizations of big sparse tensors. {\it Journal of the American Statistical Association}, 110, 1562-1576. 

\item Ziegler, A., K\"{o}nig, I.R. (2010). {\it A statistical approach to genetic epidemiology. Concepts and applications.} Weinheim:Wiley-VCH Verlag GmbH \& Co
\end{list}

\newpage
\pagebreak

\begin{figure*}[!h]
\begin{minipage}[t]{7.05cm}
{\bf \centerline{
\includegraphics[height=65mm,width=1.15\textwidth]{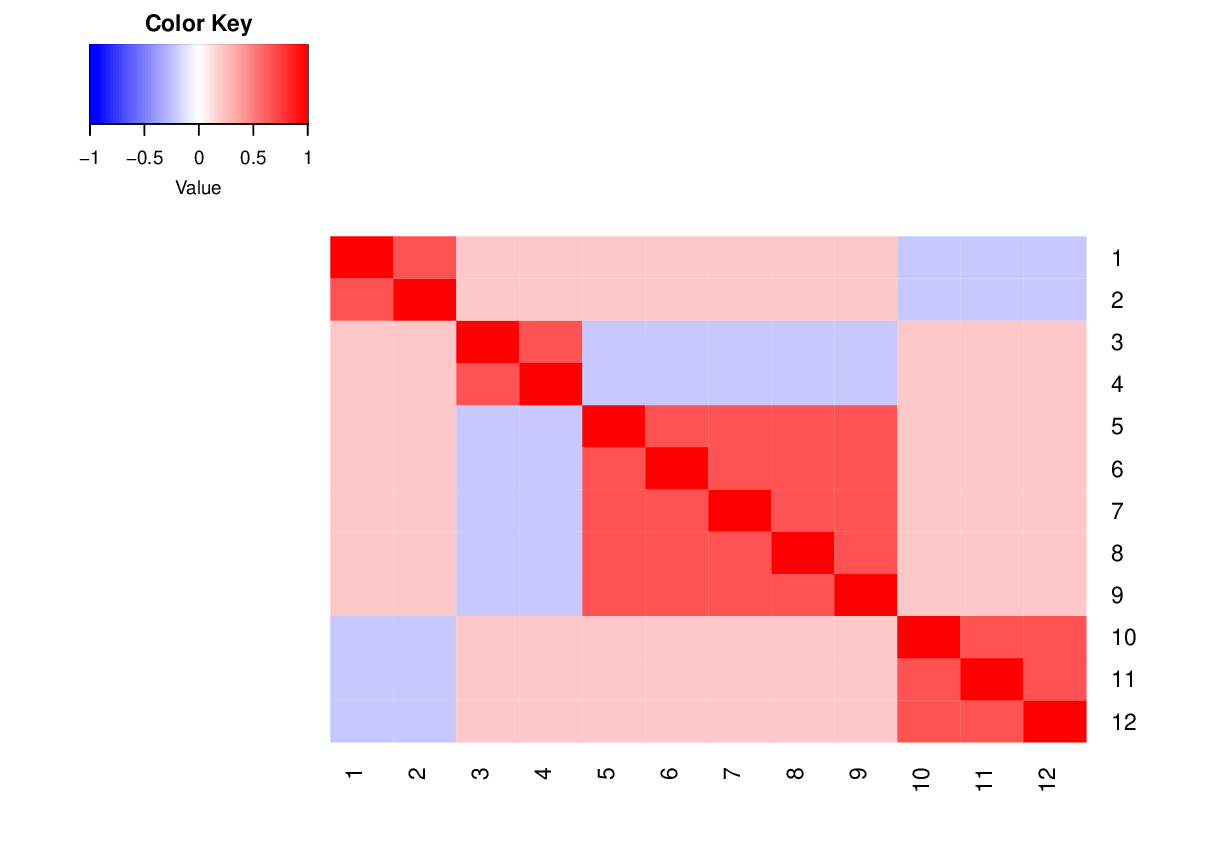}}}
\caption*{ \footnotesize{Figure 1(a): Example 2. Theoretical correlations.}}
\label{pcaegp1}
\hfill
\end{minipage}
\begin{minipage}[t]{7.05cm}
{\bf \centerline{
\includegraphics[height=65mm,width=1.15\textwidth]{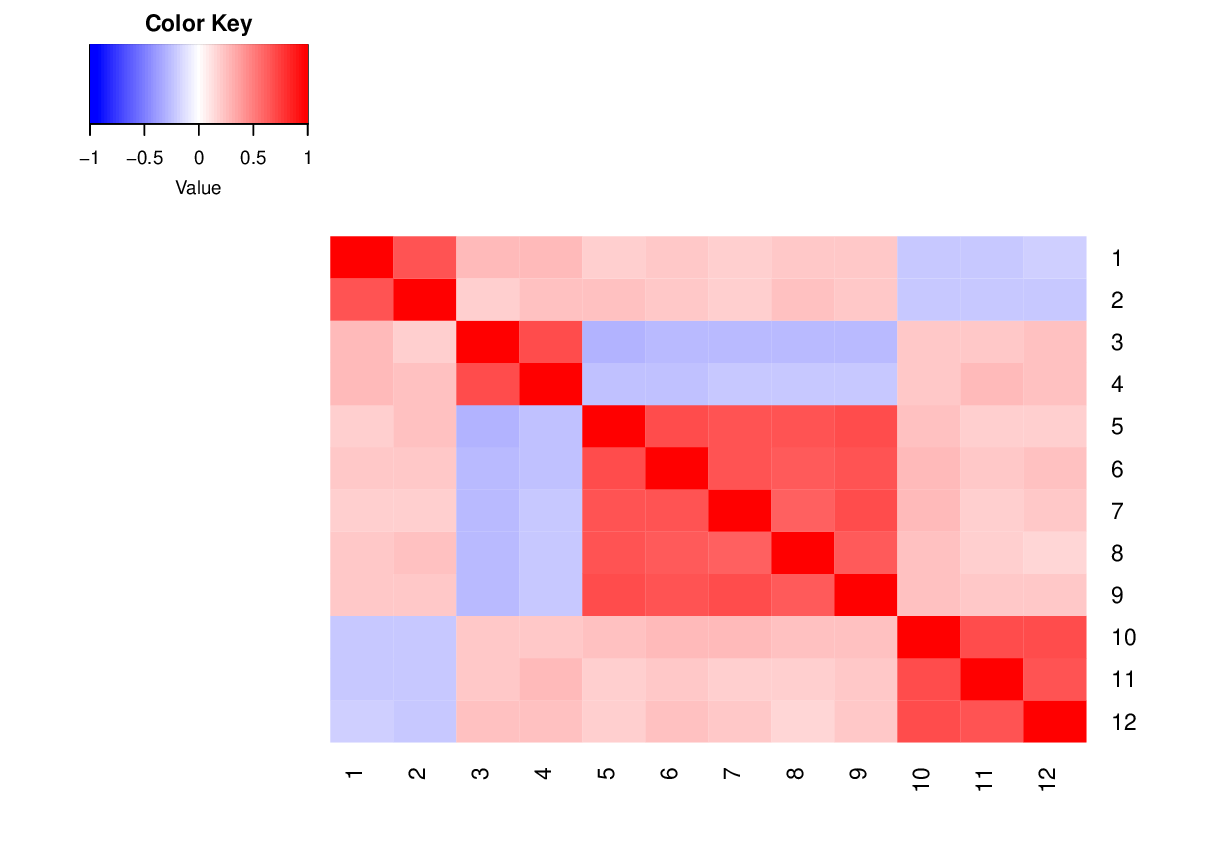}}}
\caption*{ \footnotesize{Figure 1(b): Example 2. Sample correlations.}}
\label{pcaegp2}
\end{minipage}
\end{figure*}

\begin{figure*}[!h]
\begin{minipage}[t]{7.05cm}
{\bf \centerline{
\includegraphics[height=65mm,width=1.15\textwidth]{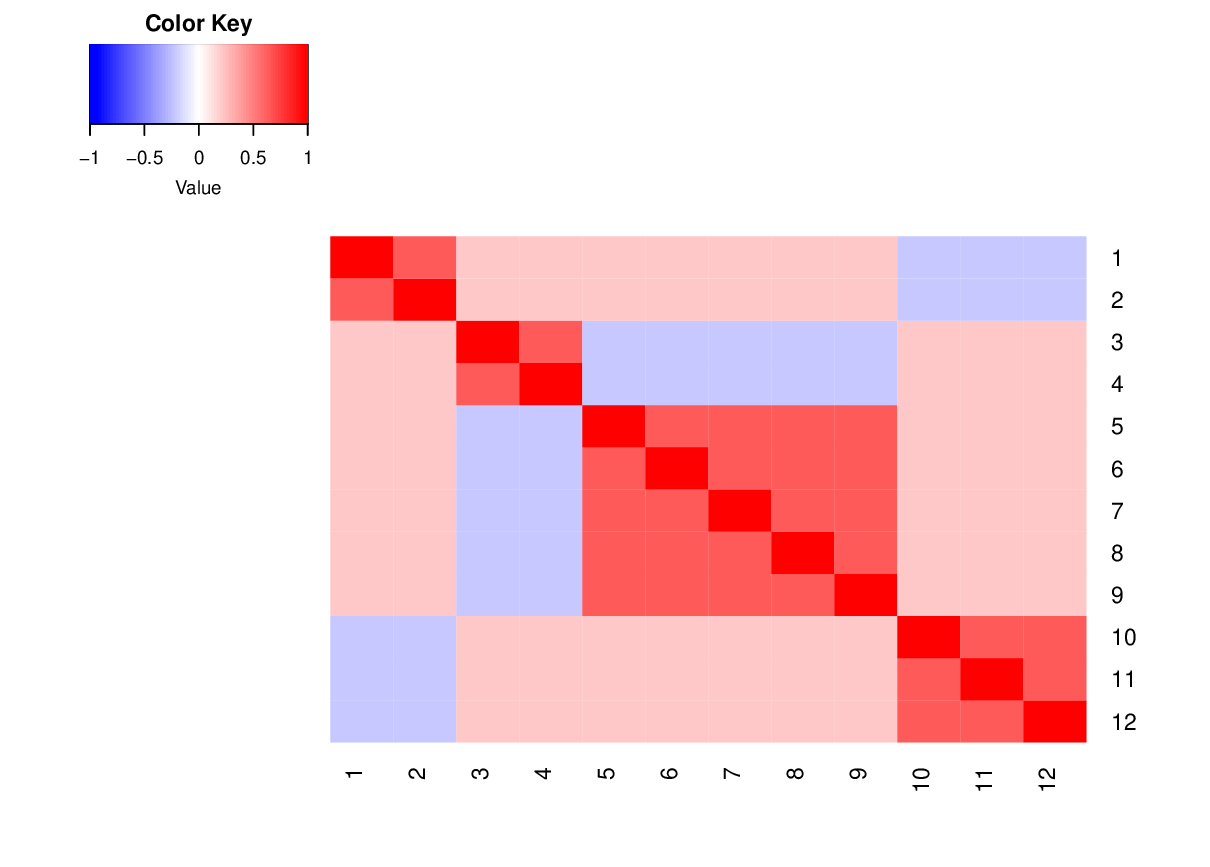}}}
\caption*{ \footnotesize{Figure 2(a): Example 3. Theoretical correlations.}}
\label{pcaegp1}
\hfill
\end{minipage}
\begin{minipage}[t]{7.05cm}
{\bf \centerline{
\includegraphics[height=65mm,width=1.15\textwidth]{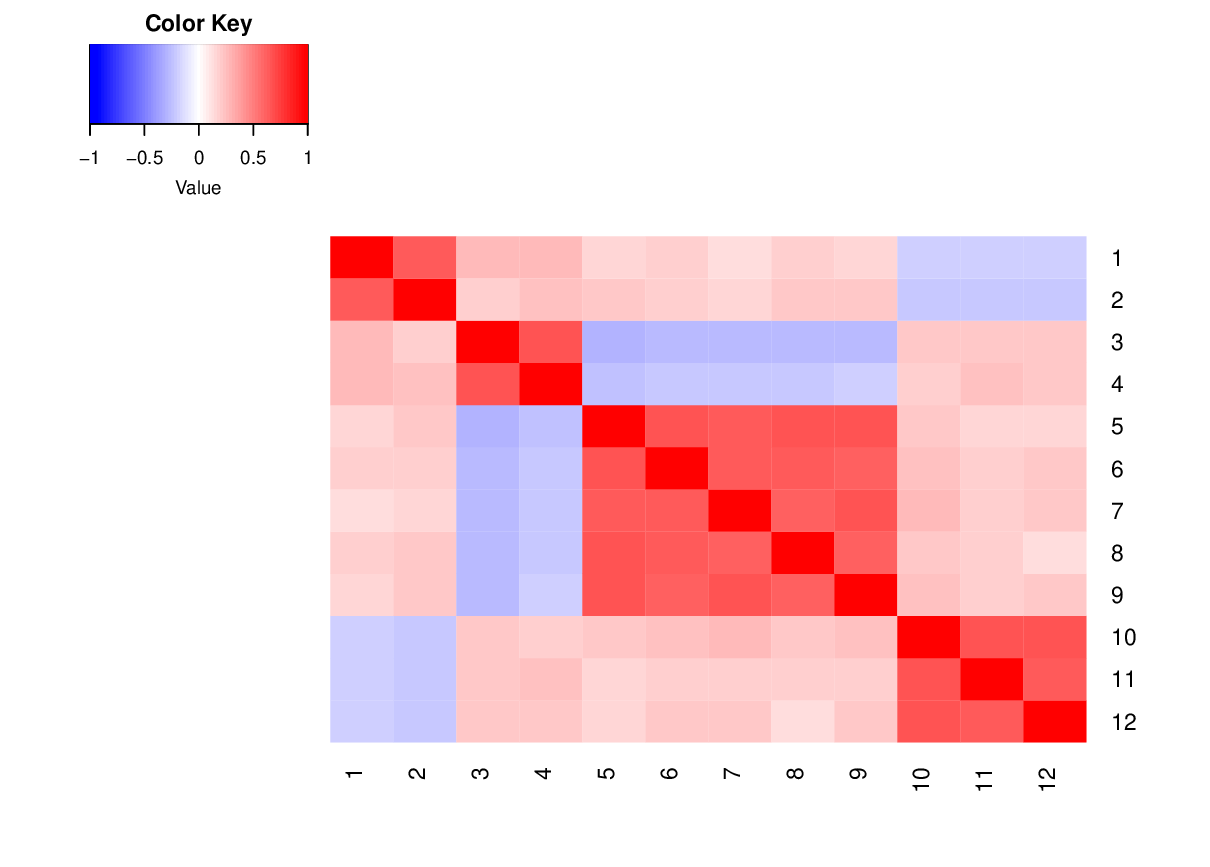}}}
\caption*{ \footnotesize{Figure 2(b): Example 3. Sample correlations.}}
\label{pcaegp2}
\end{minipage}
\end{figure*}

\begin{figure*}[!h]
\begin{minipage}[t]{7.05cm}
{\bf \centerline{
\includegraphics[height=65mm,width=1.15\textwidth]{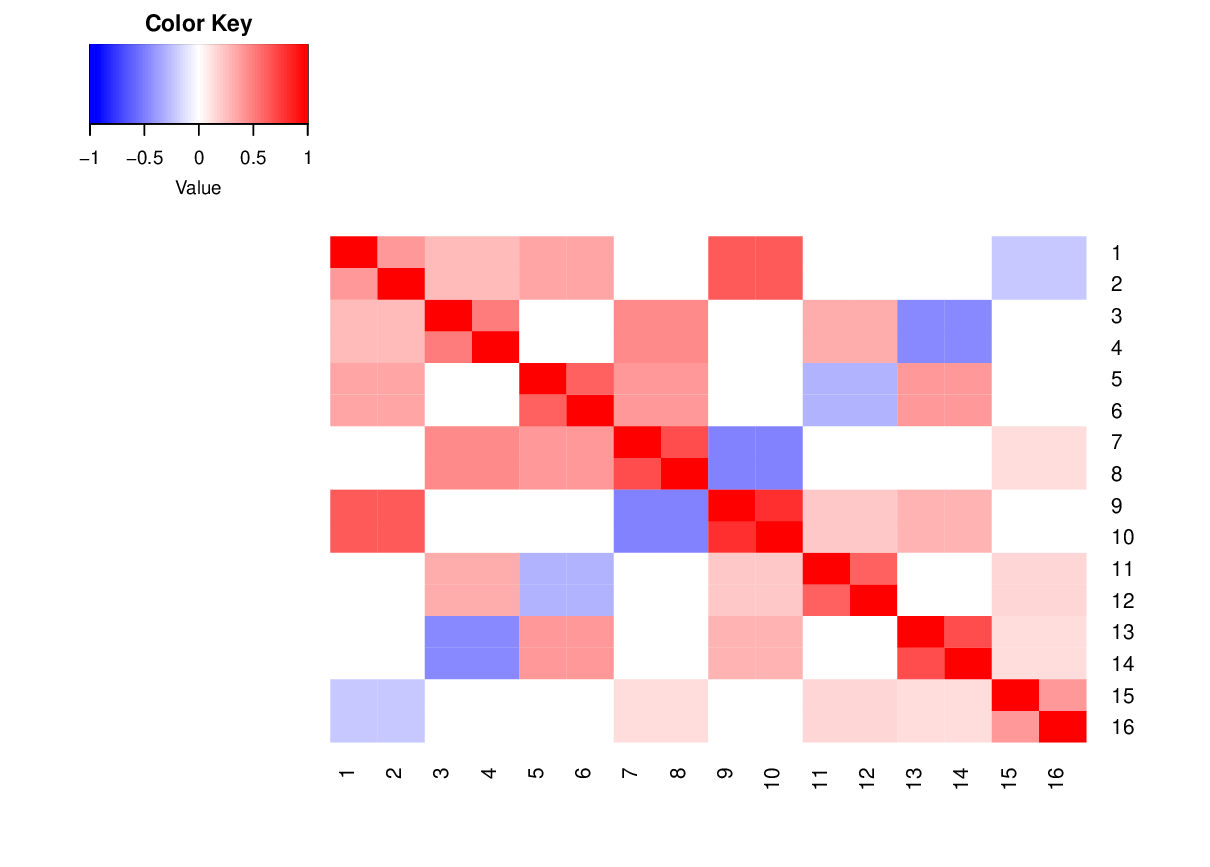}}}
\caption*{\footnotesize{Figure 3(a): Example 4. Theoretical correlations.}}
\label{pcaegp1}
\hfill
\end{minipage}
\begin{minipage}[t]{7.05cm}
{\bf \centerline{
\includegraphics[height=65mm,width=1.15\textwidth]{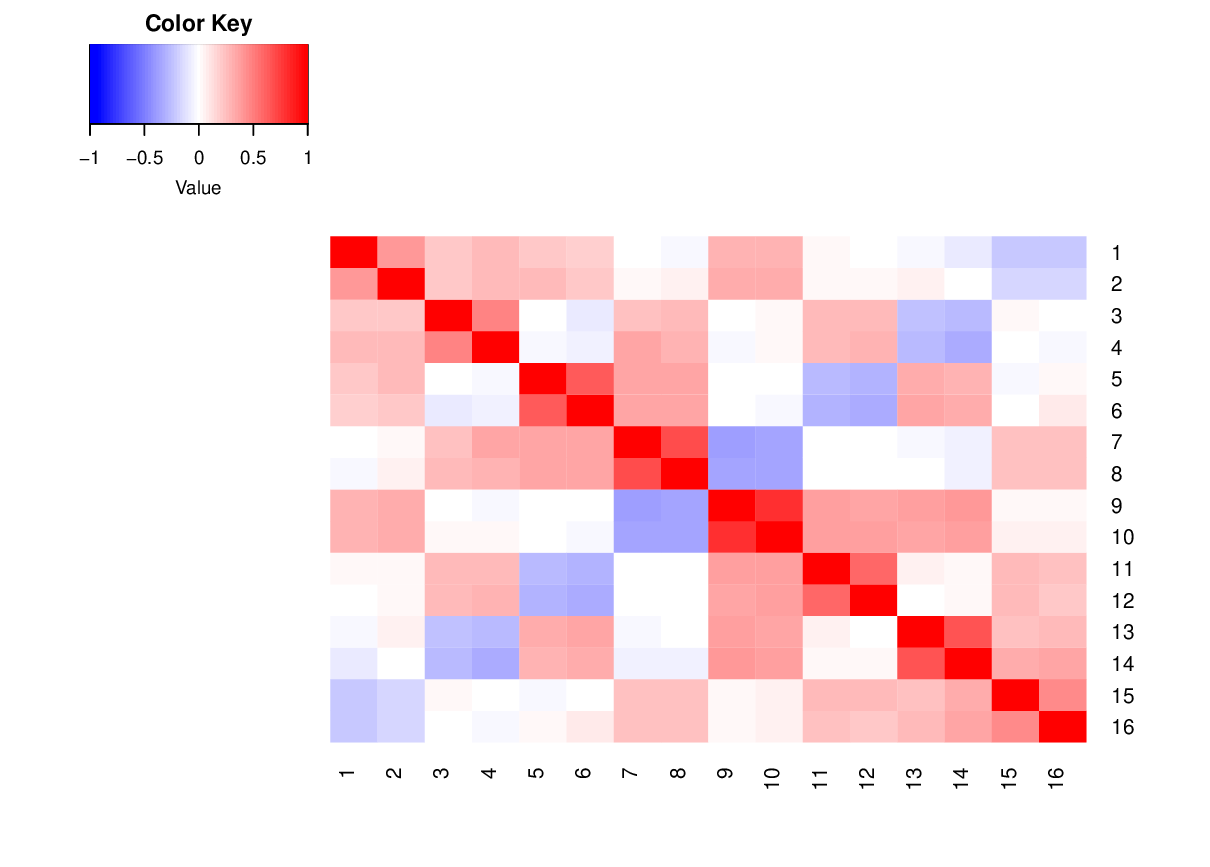}}}
\caption*{\footnotesize{Figure 3(b): Example 4. Sample correlations.}}
\label{pcaegp2}
\end{minipage}
\end{figure*}

\begin{figure*}[!h]
\begin{minipage}[t]{7.05cm}
{\bf \centerline{
\includegraphics[height=65mm,width=1.15\textwidth]{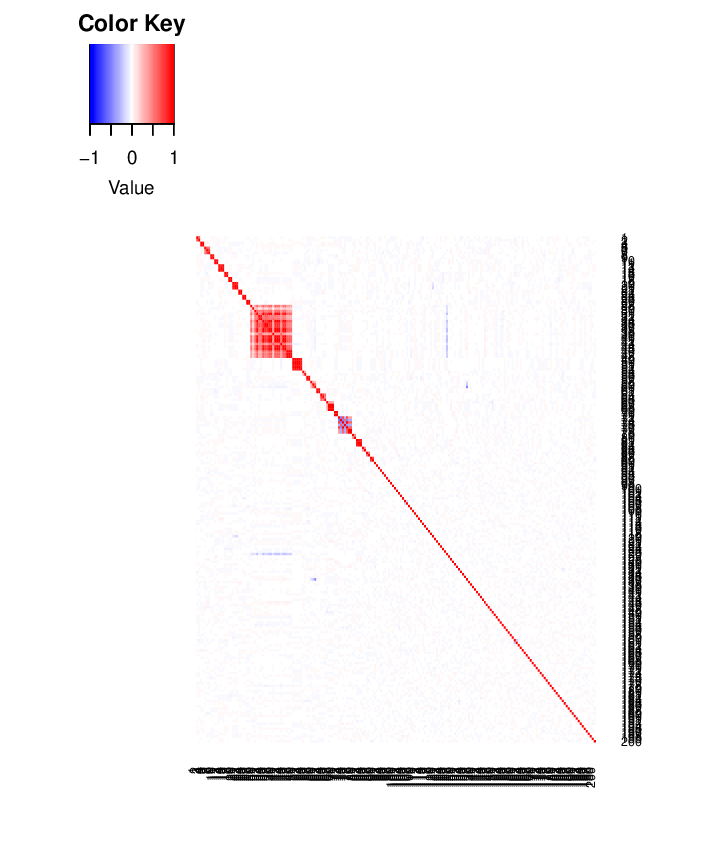}}}
\caption*{\footnotesize{Figure 4(a): Real data Linkage Disequilibrium.}}
\label{pcaegp1}
\hfill
\end{minipage}
\begin{minipage}[t]{7.05cm}
{\bf \centerline{
\includegraphics[height=65mm,width=1.15\textwidth]{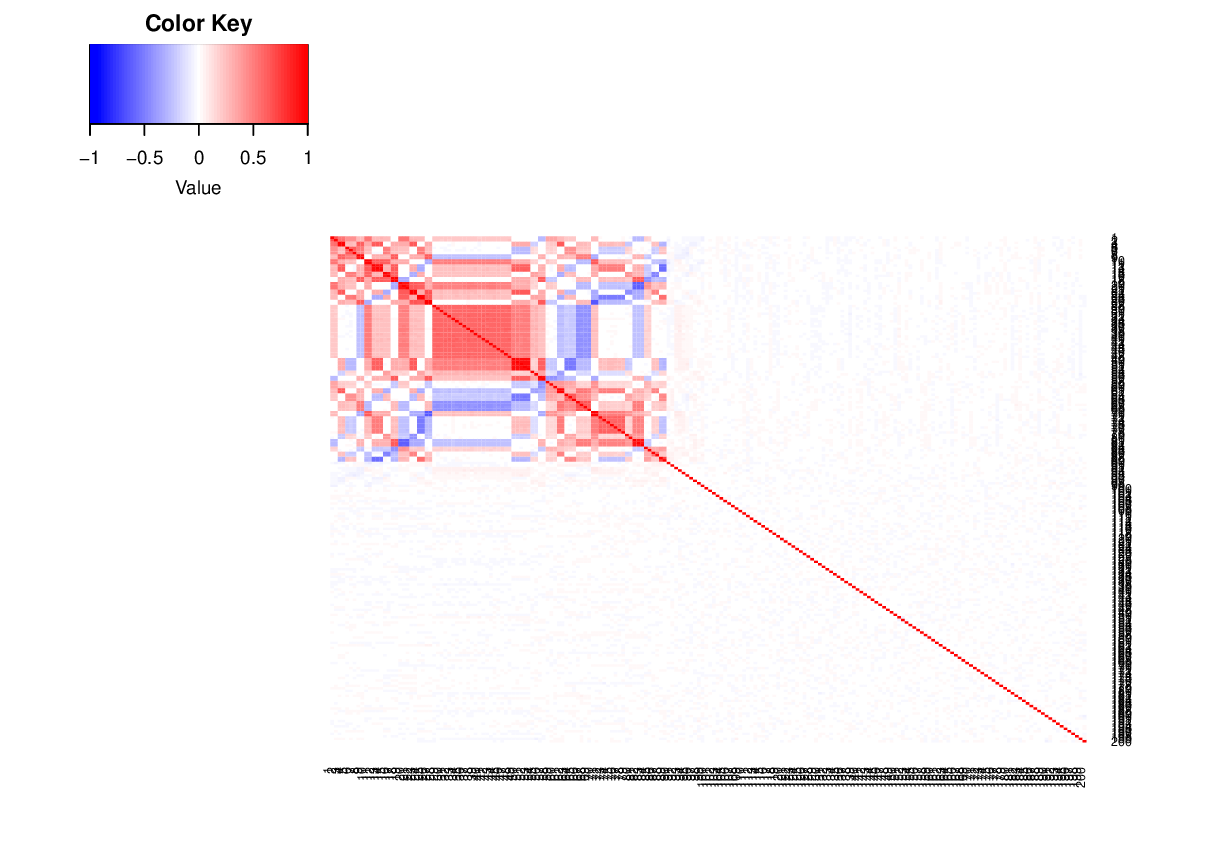}}}
\caption*{\footnotesize{Figure 4(b): Simulated data Linkage Disequilibrium.}}
\label{pcaegp2}
\end{minipage}
\end{figure*}

\end{document}